\documentclass{article}
\usepackage{ltwol2e}

\usepackage{epsfig}

\def\Journal#1#2#3#4{{#1} {\bf  #2}, #3 (#4)}

\def\NPB{{\em Nucl. Phys.} B}
\def\PLB{{\em Phys. Lett.} B}
\def\PRL{\em Phys. Rev. Lett.}
\def\PRD{{\em Phys. Rev.} D}
\def\ZPC{{\em Z. Phys.} C }
\def\EPC{{\em E. Phys. J.} C}

\def\bea{\begin{eqnarray}}
\def\eea{\end{eqnarray}}
\def\gg{\gamma\gamma}
\def\ppbar{\overline{\mbox p}\mbox{p}}

\def\sqee{\sqrt{s}_{\rm ee}}
\def\sigmagg{\sigma_{\gg}}
\def\xg{x_{\gamma}}
\def\xgp{x_{\gamma}^+}
\def\xgm{x_{\gamma}^-}
\def\xgpm{x_{\gamma}^{\pm}}
\def\etajet{\eta^{\rm jet}}

\def\qqbar{\mbox{q}\overline{\mbox{q}}}
\def\ee{\mbox{e}^+\mbox{e}^-}
\def\cost{\cos\theta^*}

\def\ETJET{E^{\rm jet}_T}
\def\pt{p_{\rm T}}
\def\dspt{{\rm d}\sigma/{\rm d}p_{\rm T}} 
\def\dsy{{\rm d}\sigma/{\rm d}y} 
\def\PTMIA{p_{\rm t}^{\rm mi}}
\def\ETBAR{\bar{E}^{\rm jet}_{\rm T}}

\begin{document}

\pagestyle{myheadings}
\markboth{FREIBURG-EHEP-98-06}{FREIBURG-EHEP-98-06}

\setcounter{footnote}{0}
\renewcommand{\thefootnote}{\fnsymbol{footnote}}
\title{FINAL STATES IN \boldmath $\gamma\gamma$ AND $\gamma$p INTERACTIONS
\footnotemark[1]
\unboldmath}

\author{STEFAN S\"OLDNER-REMBOLD}

\address{Albert-Ludwigs-Universit\"at Freiburg, 
Fakult\"at f\"ur Physik, D-79104 Freiburg, Germany\\E-mail:
stefan.soldner-rembold@cern.ch}


\twocolumn[\maketitle\abstracts{The total hadronic $\gg$ 
cross-section measured by L3 and OPAL and
the apparent discrepancy between the results are discussed.
OPAL measurements of jet and charged hadron production in
$\gg$ scattering and preliminary H1 results on $\pi^0$ production 
in $\gamma$p scattering are also presented.
The mechanism of baryon number transfer in $\gamma$p interactions at HERA
has been studied for the first time by H1.
}]

\section{Total Hadronic \boldmath $\gg$ \unboldmath Cross-Section}
\label{sec-total}
At high $\gg$ centre-of-mass energies $W=\sqrt{s}_{\gg}$, 
the total hadronic cross-section $\sigmagg$
for the production of hadrons in the interaction of 
two real photons is expected to be dominated by interactions
where the photon has fluctuated into a hadronic state. 
Measuring the $W$ dependence of $\sigmagg$ 
should therefore improve our understanding of the hadronic nature of
the photon and the universal high energy behaviour of total 
hadronic cross-sections.
\begin{figure}[htbp]
\begin{center}
\epsfig{file=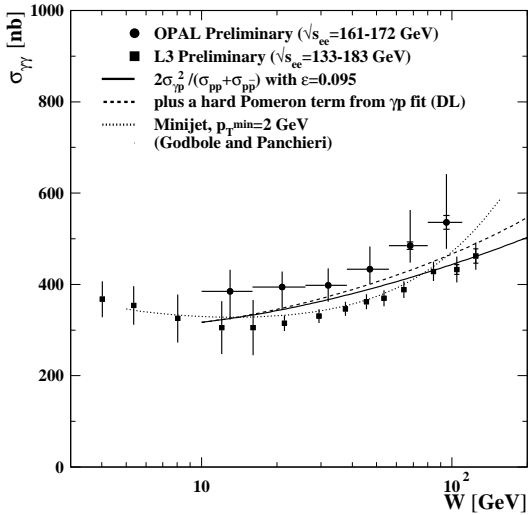,width=0.38\textwidth}
\end{center}
\caption{\label{fig-stot}
Total cross-section of the process $\gg\rightarrow\mbox{hadrons}$
as a function of $W=\sqrt{s}_{\gg}$. The inner error bars show the
statistical errors and the outer error bars the quadratic sum
of the statistical and systematic errors.}
\end{figure}

Before LEP2 the energy dependence of $\sigma_{\gg}(W)$ had only been measured
in the low energy region $W<20$~GeV~\cite{bib-oldtot}. 
These energies are too low to observe the high energy rise which is typical
for hadronic cross-sections. Using the LEP2 data,
L3~\cite{bib-l3tot} and OPAL~\cite{bib-opaltot} have now 
measured $\sigma_{\gg}(W)$ in the ranges $5 \le W \le 145$~GeV
and $10\le W \le 110$~GeV, respectively.
The results are shown in Fig.~\ref{fig-stot}.
\setcounter{footnote}{0}
\renewcommand{\thefootnote}{\fnsymbol{footnote}}
\footnotetext[1]{Talk given at ICHEP'98,
Vancouver, Canada, July 22-30, 1998}

Before interpreting these results,
the apparent discrepancy between the OPAL and L3 measurements 
and the interpretation of the systematic errors require
some discussion.
OPAL shows the average of the results obtained by
determining the detector corrections using either the Monte Carlo
model PHOJET or PYTHIA
\begin{figure}[htbp]
\begin{center}
\epsfig{file=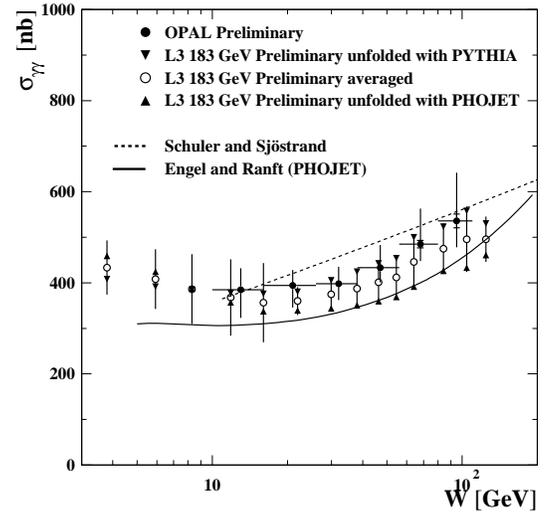,width=0.38\textwidth} 
\end{center}
\caption{\label{fig-stot2}
Comparison of the OPAL measurement with the
average of the L3 results at $\sqee=183$~GeV obtained using PHOJET
and PYTHIA (see text).}
\end{figure}
and takes the difference between these results as part of the systematic error.
L3 uses PHOJET to determine the central values 
and gives no model dependent systematic error. For $W<20$~GeV the
systematic errors become large in the case of L3 due
to the finite trigger acceptance. OPAL avoids these regions
by applying harder selection cuts. It should also be noted
that the errors are highly correlated within one experiment.
In order to make the two experiments more comparable, the
L3 results for $\sqee=183$~GeV 
obtained with the PYTHIA and the PHOJET simulation
of the detector, separately, are averaged in Fig.~\ref{fig-stot2} and
the full range is shown as vertical error bar. Comparing these
results with the OPAL measurement shows that
no significant discrepancy exists.

In Figs.~\ref{fig-stot} and \ref{fig-stot2}, 
the results are compared to various
theoretical predictions. 
The energy dependence of
the total cross-sections for $\gamma$p and hadron-hadron
collisions is well described by a Regge parametrisation of the form
\begin{eqnarray}
\sigma_{\rm AB}(s)& = & X_{\rm AB} s^{\epsilon}+Y_{1\rm AB} s^{-\eta_1}
+Y_{2\rm AB} s^{-\eta_2}  \nonumber\\
\sigma_{\rm \bar{A}B}(s)& = & X_{\rm AB} s^{\epsilon}+Y_{1\rm AB} s^{-\eta_1}
-Y_{2\rm AB} s^{-\eta_2}
\label{eq-tot1}
\end{eqnarray}
where $\sqrt{s}$ is the centre-of-mass energy of the hadron-hadron
or $\gamma$p interaction and $A,B$ denotes
the interacting particles. The first term in the equation
is due to Pomeron exchange and the other terms
are due to C-even and C-odd Reggeon exchange~\cite{bib-pdg98}. The factors
$\epsilon$, $\eta_1$ and $\eta_2$ are assumed to
be universal, whereas $X_{\rm AB}$ and $Y_{i\rm AB}$ are process dependent.

In a simple model, 
assuming factorisation of the Pomeron term $X_{\rm AB}$, 
the total $\gg$ cross-section can be related
to the $\gamma$p, pp and $\ppbar$ total cross-sections at 
high centre-of-mass energies 
$\sqrt{s}_{\gg}=\sqrt{s}_{\rm \gamma p}=\sqrt{s}_{\rm pp}=\sqrt{s}_{\rm \overline{p}p}$, where the Pomeron trajectory
should dominate:
\begin{equation}
\sigma_{\gg}
=\frac{2\left(\sigma_{\gamma{\rm p}}\right)^2}{
\sigma_{\rm pp }+\sigma_{\rm \overline{p}p}}.
\label{eq-tot2}
\end{equation}
The $\gamma$p, pp and $\ppbar$ total cross-section are
parametrised using Eq.~\ref{eq-tot1}.
The universal factor $\epsilon = 0.095\pm0.002$ predicts a 
slow rise of the total
cross-section with energy. 
Most models for the high energy
behaviour of $\sigmagg$ are based on similar factorisation assumptions
for the soft part of the cross-section. 
This ansatz gives a reasonable description of the data 
(Fig.~\ref{fig-stot}). 

More sophisticated models predict a faster rise of $\sigma_{\gg}(s)$ compared 
to the rise observed in hadron-hadron and $\gamma$p interactions
due to an additional `hard' component in photon interactions.
Donnachie and Landshoff (DL) propose an additional hard Pomeron 
term $\propto s^{0.4}$. In Fig.~\ref{fig-stot} the influence
of this term is shown, if it had the same relative size
as in $\gamma$p scattering~\cite{bib-hardDL}. 
A mini-jet model, where the rise is driven
by the increasing mini-jet cross-section~\cite{bib-minijet}, is also shown.
A faster rise is also predicted by the models of 
Schuler and Sj\"ostrand~\cite{bib-GSTSZP73} and
the model of Engel and Ranft~\cite{bib-phojet} (Fig.~\ref{fig-stot2}).
This faster rise is in qualitative agreement with the data.
Fitting a Regge parametrisation to the data,  
L3 obtains $\epsilon=0.158\pm0.006(\mbox{stat})\pm0.028(\mbox{sys})$ 
which is more than a factor 1.5 
larger than the universal value~\cite{bib-l3tot}.

Based on HERA measurements and assuming factorisation, quasi-elastic
scattering ($\gg\to\rho\rho$) and diffractive scattering
(like $\gg\to\rho X$) should contribute about $20-30\%$ to
the total hadronic $\gg$ cross-section. Most of the final
\begin{figure}[htbp]
\begin{center}
\begin{tabular}{cc}
\epsfig{file=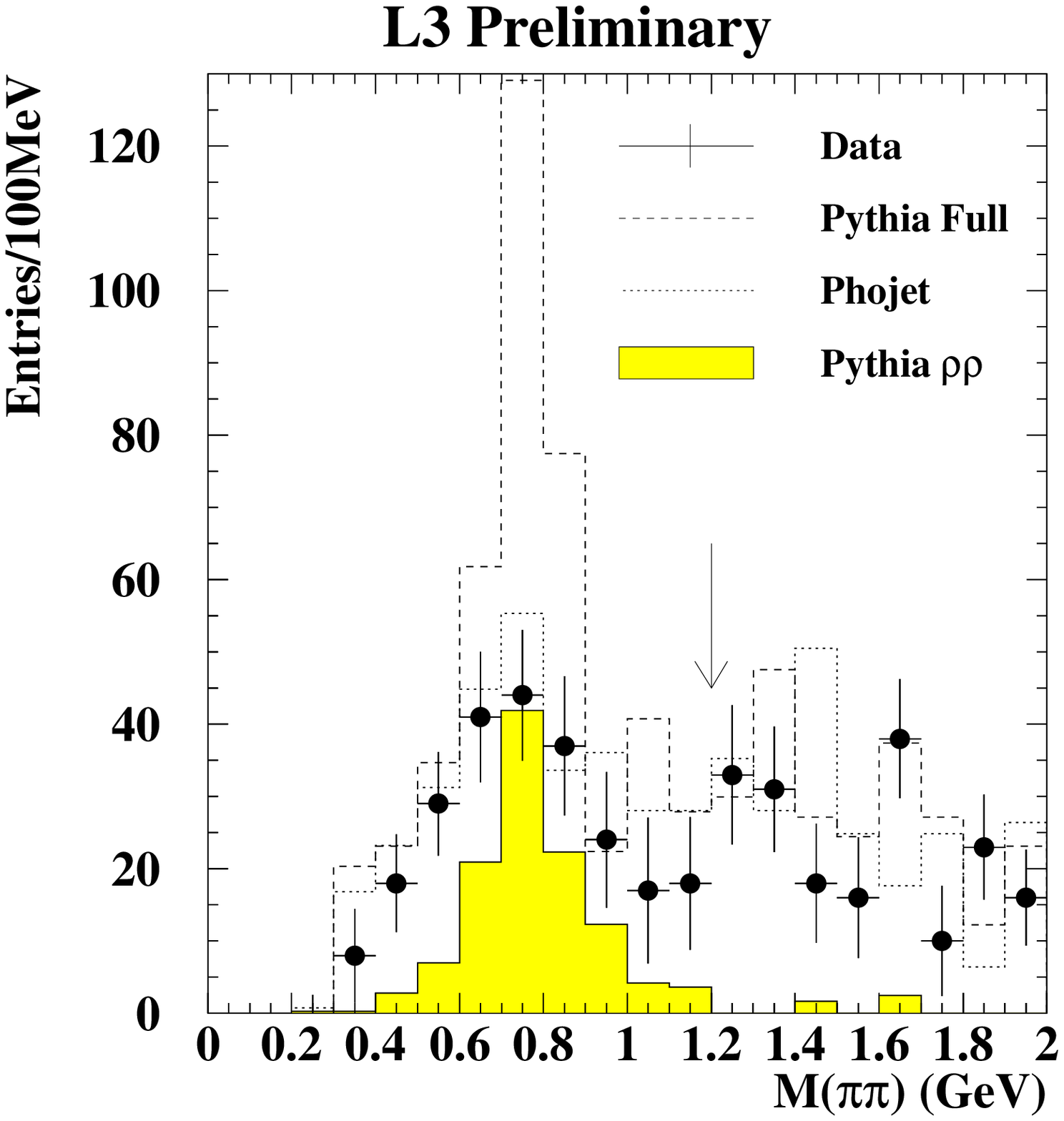,width=0.24\textwidth} &
\epsfig{file=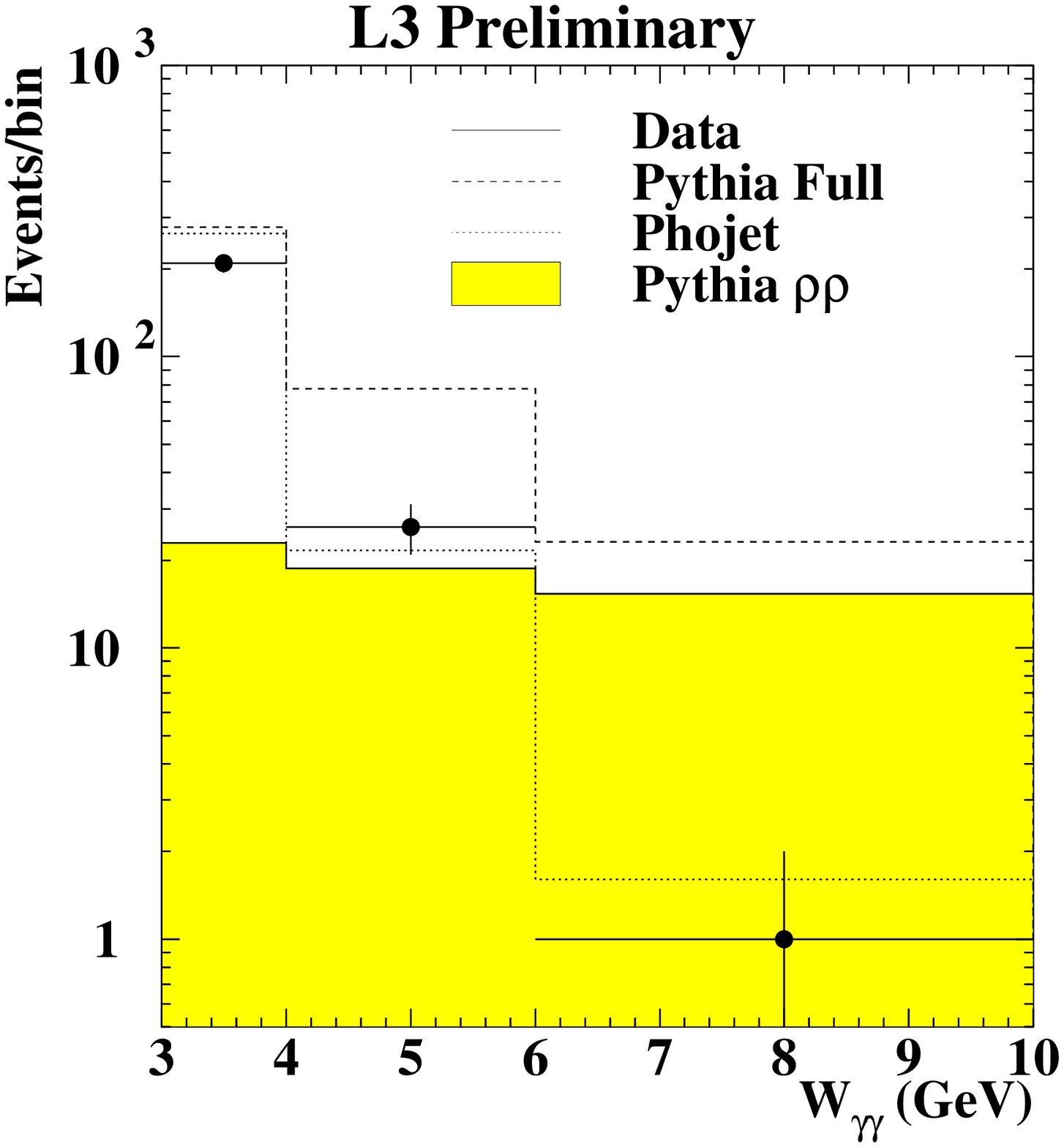,width=0.2\textwidth}
\end{tabular}
\end{center}
\label{fig-4pi}
\caption{a) $\pi^+\pi^-$ mass distribution after background subtraction
using wrong sign combination; 
b) number of $4\pi$ events as a function of $W=W_{\gg}$.}
\end{figure}
state hadrons in these processes are produced at very small
polar angles. OPAL gives an acceptance between about $6\%$ (using
PYTHIA) and about $20\%$ (using PHOJET) for the sum
of quasi-elastic and diffractive events.
In particular the acceptance for the quasi-elastic process 
$\gg\to\rho\rho$ with all 4 pions in the central tracking detectors is
extremely small at high $W$.

L3 has measured the rate of $\gg\to\rho\rho$ events in the
range $3<W<10$~GeV. The
$\pi^+\pi^-$ mass distribution after background subtraction
using wrong sign combinations and the
number of $4\pi$ events as a function of $W$ are shown in Fig.~\ref{fig-4pi}.
The distribution is compared to the full PYTHIA and PHOJET simulation
and to the PYTHIA simulation of $\gg\to\rho\rho$ events. Both
models do not contain any $\gg\to{\mbox \em Resonance}$ processes
which should also contribute in this mass region. From this study
L3 concludes that both the rate and the $W$ dependence of 
quasi-elastic processes are not properly simulated by PYTHIA.
\section{Leading Order Parton Processes}
The interaction of quasi-real photons ($Q^2\approx0$) studied at LEP and
the interaction of a quasi-real photon with a proton studied
at HERA (photoproduction)
are very similar processes. 
In leading order (LO) different event classes can be defined
in $\gg$ and $\gamma$p interactions. The photons can either
interact as `bare' photons (``direct'') or as hadronic fluctuation
(``resolved''). 
Direct and resolved interactions can be separated by measuring
the fraction $\xg$ of the photon's momentum participating in the 
hard interaction for the two photons. In $\gg$ interactions
they are labelled $\xgpm$ for the two photons.
Ideally, the direct $\gg$ events with two bare
photons are expected to have $\xgp=1$ and
$\xgm=1$,
whereas for double-resolved events both values $\xgp$ and $\xgm$
are expected to be much smaller than one. 
In photoproduction, the
interaction of a bare photon with the proton is labelled `direct'
(corresponding to the `single resolved' process in $\gg$) and
the interaction of a hadronic photon is called `resolved'
(corresponding to `double-resolved' in $\gg$).

\section{Di-Jet Production in \boldmath $\gg$ \unboldmath Interactions}
\label{sec-jet}
Studying jets should give access to the parton dynamics
of $\gg$ and $\gamma$p interactions.
OPAL has therefore measured di-jet production in $\gg$ scattering
at $\sqee=161-172$~GeV using the cone jet finding algorithm with 
$R=1$~\cite{bib-opaljet}. 
Similar studies have been presented by the HERA 
experiments~\cite{bib-hera}.
In di-jet events, $\xgpm$ is calculated using
\begin{equation}
\xgp=\frac{\displaystyle{\sum_{\rm jets=1,2}(E+p_z)}}
 {{\displaystyle\sum_{\rm hadrons}(E+p_z)}} \;\;\;\mbox{and}\;\;\;
\xgm=\frac{\displaystyle{\sum_{\rm jets=1,2}(E-p_z)}}
{\displaystyle{\sum_{\rm hadrons}(E-p_z)}},
\nonumber
\label{eq-xgpm}
\end{equation}
where $p_z$ is the momentum component along the $z$ axis of the
detector and $E$ is the energy of the jets or hadrons.

For a given jet-jet centre-of-mass energy the cross-sections
vary only with the scattering angle $\theta^{*}$.
The leading order direct process $\gg \rightarrow \qqbar$ is mediated
by $t$-channel spin-$\frac{1}{2}$ 
quark exchange which leads to an angular dependence
$\propto (1-\left|\cos\theta^{*2}\right|)^{-1}$.
\begin{figure}[htbp]
\begin{center}
\epsfig{file=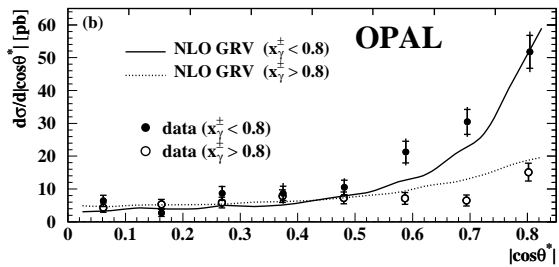,
width=0.4\textwidth}
\end{center}
\caption{Angular distribution of di-jet events
compared to NLO QCD calculations
using the GRV parametrisation.
The curves are normalised to the data in the first
three bins. The invariant mass of the two-jet system
must be larger than 12 GeV.}
\label{fig-theta}
\end{figure}
In double-resolved processes all matrix elements
involving quarks and gluons have to be taken into account,
with a large contribution from spin-$0$ gluon exchange. 
After adding up all relevant processes, perturbative QCD
predicts an angular dependence of approximately $\propto
(1-\left|\cost\right|)^{-2}$.
Fig.~\ref{fig-theta} shows the corrected $|\cost|$ distribution
of di-jet events with $\xgpm>0.8$ and with $\xgpm < 0.8$
compared to a NLO calculation~\cite{bib-klasen2} which 
qualitatively reproduces the data.

The transverse momentum $\ETJET$
of the jet (or the final-state parton) defines a hard scale
which can be used together with $\xg$ to constrain the
parton densities $f(\xg,E^2_{\rm T})$ of the photon.
In the kinematic range covered by LEP
the $F_2^{\gamma}$ measurements are mainly probing
the quark content of the photon~\cite{bib-klaus}, whereas
di-jet production can be used
to constrain the relatively unknown gluon distribution in the photon.

\begin{figure}[htbp]
\begin{center}
\epsfig{file=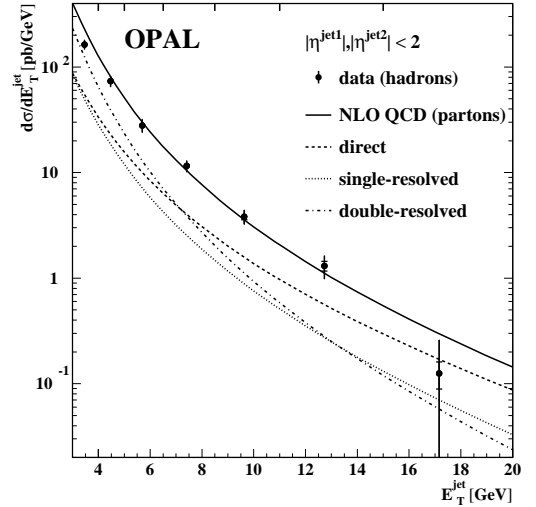,
width=0.37\textwidth}
\end{center}
\caption{\label{fig-ettwojet}
The inclusive $\ee$ di-jet cross-section as a function
of $\ETJET$ for jets with $|\etajet|<2$ using a cone size $R=1$.}
\end{figure}

The $\ETJET$ distribution for di-jet events with pseudorapidities 
$|\etajet|<2$ 
is shown in Fig.~\ref{fig-ettwojet}. The measurements are
compared to a NLO calculation~\cite{bib-kleinwort} 
which uses the NLO GRV parametrisation~\cite{bib-grv}.
The direct, single- and double-resolved parts and their sum are
shown separately. The data points are in good agreement with
the calculation except in the first bin where
theoretical and experimental uncertainties are large.

\begin{figure}[htbp]
\begin{center}
\epsfig{file=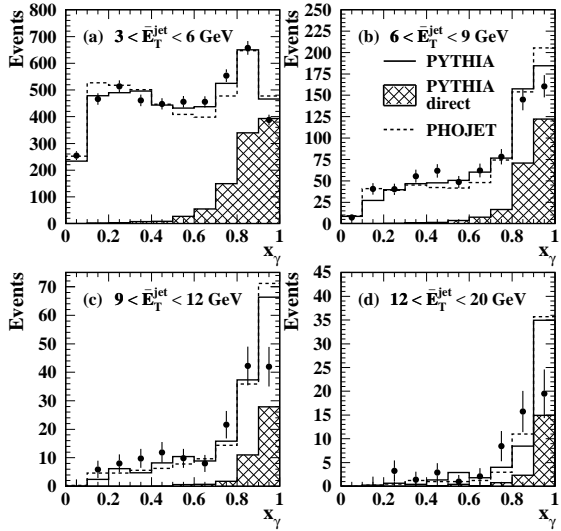,width=0.4\textwidth} 
\end{center}
\caption{Uncorrected $\xg$ distribution 
in bins of the mean value of 
$\ETBAR$ for di-jet events with $\ETJET>3$~GeV and $|\etajet|<2$. 
Each event is added
to the plot twice, at the values of $\xgp$ and of $\xgm$.
Statistical errors only are shown.}
\label{fig-xg}
\end{figure}

The $\xg$ distribution is shown in Fig.~\ref{fig-xg} in bins of $\ETBAR$,
where $\ETBAR$ is the mean value of the transverse energies 
of the two jets. No detector correction has been applied.
The Monte Carlo predictions of PYTHIA and PHOJET are normalised
to the number of data events.
The contribution from direct processes, as predicted from PYTHIA,
is also shown. The events from direct processes are concentrated
at high $\xg$ values.
As $\ETBAR$ increases, the $\xg$ distribution
shifts to higher values and the fraction of direct events in the PYTHIA
sample increases.
The number of events is underestimated by PYTHIA and PHOJET
by about $25-30 \%$, if the predicted Monte Carlo cross-sections are
taken into account, mainly for $\xg<0.9$.

The NLO QCD calculations do not take into account the possibility
of an underlying event which leads to an increased
jet cross-section. The underlying event is simulated in the Monte Carlo models
PYTHIA and PHOJET which will be used to compare to
different LO parametrisations of
the parton distribution, 
GRV~\cite{bib-grv}, SaS-1D~\cite{bib-sas} and LAC1~\cite{bib-LAC1}. 
In PYTHIA and PHOJET the modelling of the
underlying event includes multiple interactions. A resolved
photon contains several partons which can lead to multiple
parton interactions in double-resolved events.

The contribution from multiple interactions has to be tuned
using quantities which are not directly correlated to the jets,
since otherwise effects of the parton distributions and
of the underlying event cannot be distinguished.

It is expected that the transverse energy flow outside the jets
measured as a function of $\xg$
is correlated to the underlying event~\cite{bib-h1}.
No effect due to the underlying event
is expected for direct events at large $\xg$.
The increase of the transverse energy flow outside
the two jets at small $\xg$ can therefore be used
to tune the number of multiple interactions in the model.

\begin{figure}[htbp]
   \begin{center}
      \mbox{
          \epsfxsize=8.5cm
          \epsffile{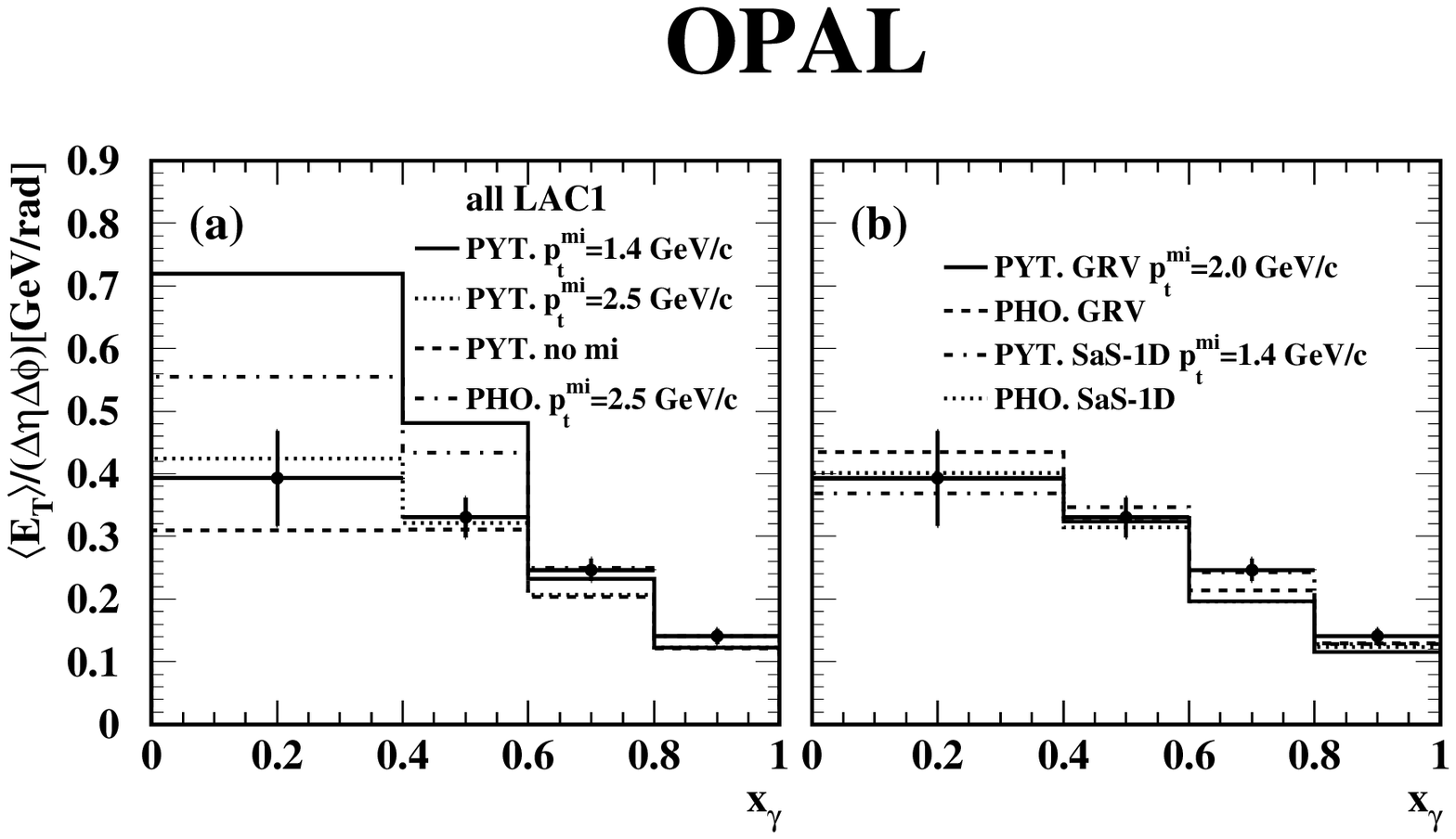}
           }
   \end{center}
\caption{Transverse energy flow outside the jets in the central
 rapidity region $|\eta^*| < 1$ as a function of $\xg$. 
 The statistical error is smaller
 than the symbol size. The error bars show the statistical
 and systematic errors added in quadrature.
 The data are compared with the MC models using a) LAC1 and b) GRV and SaS-1D.}
\label{fig-mia}
\end{figure}

The events were boosted into their
centre-of-mass system and the transverse energy flow was measured
as a function of $\xg$ in the central rapidity region $|\eta^*|<1$. 
The regions around the jet axes with $R<1.3$ are excluded from the
energy sum.
Fig.~\ref{fig-mia} shows the transverse
energy flows corrected to the hadron level compared
to the Monte Carlo models with different values
of the parameter $\PTMIA$ which defines the
transverse momentum cutoff for multiple parton interactions.
The following conclusion can be drawn:
\begin{itemize}
\item The influence of multiple interactions is small.
The modelling of the transverse energy flow without multiple
interactions is also consistent with the data.
\item
The optimised value of $\PTMIA$ depends on the parametrisation
used for the parton distributions.
For all further comparisons with PYTHIA,
the cutoff parameter $\PTMIA$ was set to $2.5$~GeV$/c$
for LAC1, to $2.0$~GeV$/c$ for GRV and to $1.4$~GeV$/c$ for SaS-1D.
\item 
PHOJET with either SaS-1D or GRV is in reasonable agreement with the data.
Changing the default cutoff from $\PTMIA= 2.5$~GeV$/c$ 
does not affect the transverse energy flow significantly.
\item For $\gamma$p collisions at HERA, $\PTMIA=1.2$~GeV$/c$ is
the optimal choice with PYTHIA-GRV
and $\PTMIA=2.0$~GeV$/c$ with PYTHIA-LAC1~\cite{bib-h1}. With these 
values the models slightly overestimate
the transverse energy flows at low $\xg$ in the $\gg$ data. 
\end{itemize}

\begin{figure}[htbp]
   \begin{center}
      \mbox{
          \epsfxsize=8.5cm
           \epsffile{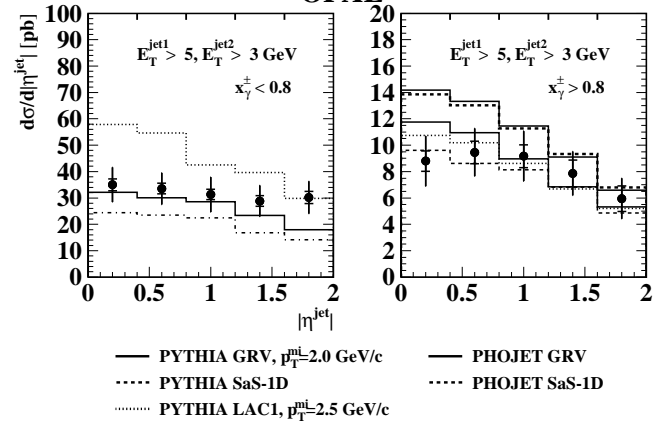}
           }
   \end{center}
\label{fig-etatwojet}
\caption{The inclusive two-jet cross-section as a function of $|\etajet|$
 for events with $E^{\rm jet1}_{\rm T}> 4$~GeV and
 $E^{\rm jet2}_{\rm T}> 3$~GeV is
 shown for events with a) $\xgpm < 0.8$ (mainly double resolved)
 and b) $\xgpm > 0.8$ (mainly direct).
 Asymmetric $\ETJET$ cuts were
 chosen to avoid singularities in the NLO calculations (not shown).}
\end{figure}
After this optimisation of the description of the underlying
event by the generators, the
sensitivity of the jet cross-sections to the choice of parametrisation
for the parton distributions can be studied.
The inclusive di-jet cross-section as a function of $|\etajet|$
for events with a large double-resolved 
contribution obtained by requiring $\xgpm<0.8$ is shown in
Fig.~\ref{fig-etatwojet}a. 
The larger gluon density in LAC1 compared to SaS-1D and
GRV in the $(\xg,\ETJET)$ region probed here leads
to an overestimation of the jet cross-section for double-resolved events.
As expected, there exists almost no dependence on the choice
of parametrisation for the mainly direct events with $\xgpm>0.8$
in Fig.~\ref{fig-etatwojet}b. 

\section{Hadron Production in \boldmath $\gg$ and $\gamma$p
\unboldmath Scattering}
Measurements of hadron production cross-sections in $\gg$
and $\gamma$p scattering complement the studies of
jet production. Hadron production at large transverse momenta
is sensitive to the partonic structure of the interactions
without the theoretical and experimental problem related
to the various jet algorithms. Interesting comparisons
of $\gg$ and $\gamma$p data taken at LEP and HERA, respectively,
should be possible in the future, since similar hadronic
centre-of-mass energies $W$ of the order 100 GeV are accessible
for both type of experiments.
\subsection{Inclusive Charged Hadron Production in $\gg$}
\begin{figure}[htbp]
   \begin{center}
      \mbox{
          \epsfxsize=0.36\textwidth
          \epsffile{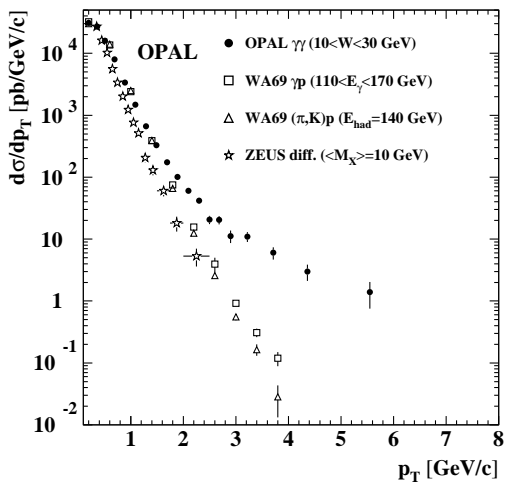}
          }
   \end{center}
\caption{\label{fig-wa69}
The $\pt$ distribution measured in $\gg$ interactions
in the range $10<W<30$~GeV is compared to the $p_{\rm T}$ distribution
measured in $\gamma$p and hp (h$=\pi,$K) interactions by 
WA69~\protect\cite{bib-wa69}. 
The cross-section values on the ordinate
are only valid for the OPAL data.}
\end{figure}

The distributions of the transverse momentum $\pt$ 
of hadrons produced in $\gg$ interactions 
are expected to be harder than in $\gamma p$ or hadron-p
interactions due to the direct component. 
This is demonstrated in Fig.~\ref{fig-wa69} by comparing 
$\dspt$ for charged hadrons measured in $\gg$ interactions by 
OPAL~\cite{bib-opalhad} to the $p_{\rm T}$ distribution 
measured in $\gamma$p and hp (h$=\pi,$K) interactions by WA69~\cite{bib-wa69}. 
The WA69 data are normalised to the $\gg$ data in the low 
$p_{\rm T}$ region 
at $\pt\approx 200$~MeV/$c$ using the same factor for the hp and the
$\gamma$p data.
The hadronic invariant
mass of the WA69 data is about $W=16$~GeV which is
of similar size as the average $\langle W \rangle$ of the $\gg$
data in the range $10<W<30$~GeV.

Whereas only a small increase is observed
in the $\gamma$p data compared to the h$\pi$ data at large $\pt$,
there is a significant increase of the relative rate in the range 
$\pt>2$~GeV/$c$ for $\gg$ interactions due to the
direct process. 
\begin{figure}[htbp]
   \begin{center}
      \mbox{
          \epsfxsize=0.3\textwidth
\epsffile{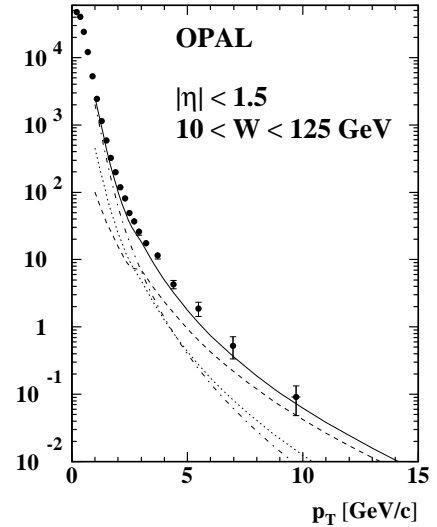}
          }
   \end{center}
\caption{\label{fig-dspt}
$\dspt$ for pseudorapidities $|\eta|<1.5$ in the range 
$10<W<125$~GeV compared to NLO calculations for $\pt>1$~GeV$/c$ 
(continuous curve) together with the double-resolved 
(dot-dashed), single-resolved (dotted)
and direct contributions (dashed).}
\end{figure}

The $\gg$ data are also compared to
a ZEUS measurement of charged particle production 
in $\gamma$p
events with a diffractively dissociated photon at $\langle W \rangle = 
180$~GeV. The invariant mass relevant for this comparison 
should be the mass $M_{\rm X}$ of the
dissociated system (the invariant mass of the `$\gamma$-Pomeron' system).
The average $\langle M_{\rm X} \rangle$ equals 10 GeV for the data shown.
The $\pt$ distribution falls exponentially, similar to the
$\gamma$p and hadron-p data, and shows no flatening
at high $\pt$ due to a possible hard component of the Pomeron.

The cross-sections $\dspt$ are also compared to NLO
calculations~\cite{bib-binnewies} which
are calculated using the QCD partonic cross-sections,
the NLO GRV parametrisation
of the parton distribution functions~\cite{bib-grv} and
fragmentation functions fitted to e$^+$e$^-$ data.
The renormalisation and factorisation scales
are set equal to $\pt$. The change in slope around $\pt=3$~GeV/$c$ in the
NLO calculation is due to the charm threshold.

In Fig.~\ref{fig-dspt}, the NLO calculation is shown 
separately for direct, single- and
double-resolved interactions. At large $\pt$ the direct 
interactions dominate. The agreement between the data and the NLO
calculation is good. 

\vspace{5mm}

\noindent{\em 4.2~~Inclusive $\pi^0$ Production in $\gamma$p}

\vspace{3mm}

\noindent
H1 has studied $\pi^0$ production in photoproduction by reconstructing
the $\pi^0\to\gg$ decays using the new
lead-scintillating fibre calorimeter SpaCal
in the backward region (photon hemisphere)~\cite{bib-h1pi0}.
\begin{figure}[htbp]
\begin{center}
\epsfig{file=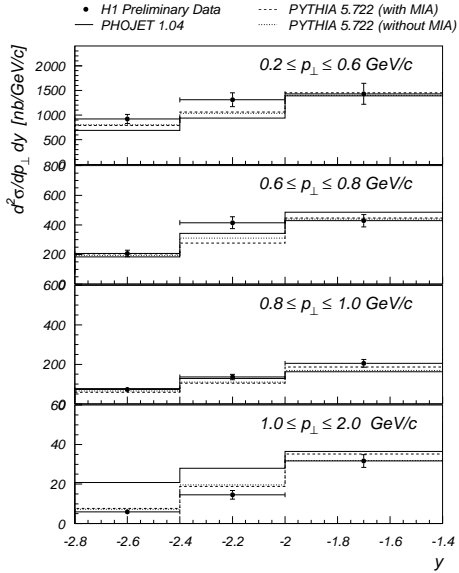,
width=0.34\textwidth}
\end{center}
\caption{\label{fig-ypi0}
$\dsy$ for $\pi^0$ produced in $\gamma$p interactions
in different transverse momentum intervals. The results
are compared to the PYTHIA and PHOJET predictions.
}
\end{figure}
Fig.~\ref{fig-ypi0} shows the $\pi^0$ cross-section as
a function of the rapidity $y$ together with the model
predictions of PHOJET and PYTHIA. PYTHIA gives a better
description of the data than PHOJET. The accuracy of the data
is, however, not yet sufficient to distinguish
multiple interaction models.

\begin{figure}[htbp]
\begin{center}
\epsfig{file=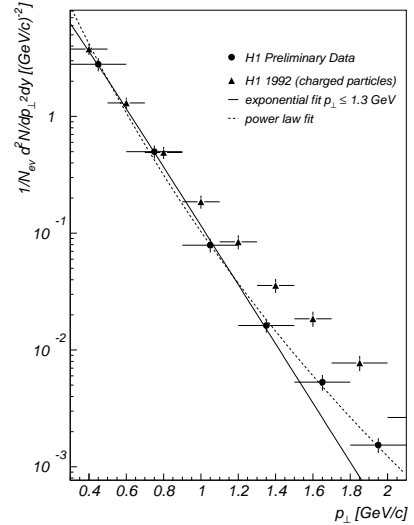,
width=0.32\textwidth}
\end{center}
\caption{\label{fig-ptpi0}
$\dspt$ for $\pi^0$ produced in $\gamma$p interactions
The triangles indicate the H1 result on charged hadrons
in the central rapidity region.
The curves are different fits to the data.}
\end{figure}
The differential cross-section $\dspt$ is presented in Fig.~\ref{fig-ptpi0}
together with H1 results on charged hadron production in
the central pseudorapidity region $(|\eta|<1.5)$.
Fitting an exponential and a power law function to the $\pt$ spectrum
shows that the low $\pt$ region, $\pt<1.3$~GeV, is well described by
an exponential fall-off typical for soft hadronic
interactions, but at high $\pt$ a deviation is observed.
In this region a power law function, which is
typical for hard scattering processes, fits the data best.
\section{Baryon-Antibaryon Asymmetry}
It has been suggested by Kopeliovich and Povh~\cite{bib-povh} that
the Baryon Number (BN) of the proton in $\gamma$p scattering 
can either be carried by
the valence quarks or by the sea quarks and gluons. The
gluonic mechanism of BN transport proceeds through the production of
baryon-antibaryon pairs in the photon fragmentation region.

\begin{figure}[htbp]
   \begin{center}
    \mbox{
          \epsfxsize=0.40\textwidth
          \epsffile{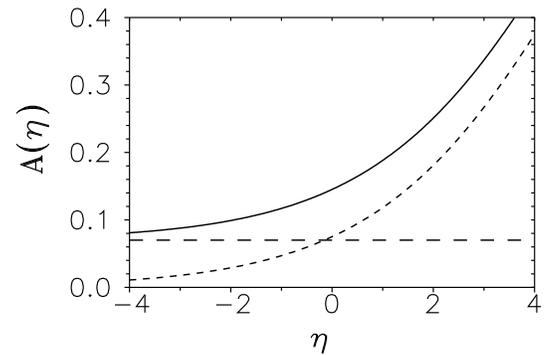}
          }
   \end{center}
\caption{\protect\label{fig-pppp}
          Baryon-antibaryon asymmetry $A_{\rm B}(\eta)$ as a function
          of rapidity $\eta$ for
          photoproduction at HERA~\protect\cite{bib-povh}.
          The dashed line corresponds to the gluonic mechanism of BN transfer
          and the dotted line to the quark mechanism which is peaked
          in the forward direction. The sum is shown as continuous line.} 
\end{figure}

H1 has studied this phenomenon in tagged photoproduction 
($150<W<260$~GeV) by measuring the baryon-antibaryon asymmetry 
\begin{equation}
A_{\rm B}
=2\frac{N_{\rm p}-N_{\rm \overline{p}}}{N_{\rm p}+N_{\rm \overline{p}}}
\end{equation}
from the number of protons, $N_{\rm p}$, and antiprotons,
$N_{\rm \overline{p}}$, per event~\cite{bib-h1p}. The theoretical expectation
for the rapidity dependence of $A_{\rm B}$ is shown in 
Fig.~\protect\ref{fig-pppp}.

In the H1 detector, the (anti-)protons are identified by
requiring that the measured energy loss, ${\rm d}E/{\rm d}x$, in
the Central Jet Chamber is twice the energy loss expected for a
minimum ionising particle. Additional cuts reduce the background
from beam-gas interactions and from secondary protons produced
in the beam pipe. The baryon-antibaryon asymmetry is
measured to be $$A_{\rm B}=0.8\pm 1.0\mbox{(stat)}\pm 2.5\mbox{(sys)}\%$$
for (anti-)protons with momenta $0.3<p<0.6$~GeV/$c$ and
for polar angles $|\cos\theta|<0.8$~$(|\eta|<1.1)$.
\begin{figure}[htbp]
   \begin{center}
    \mbox{
          \epsfxsize=0.46\textwidth
          \epsffile{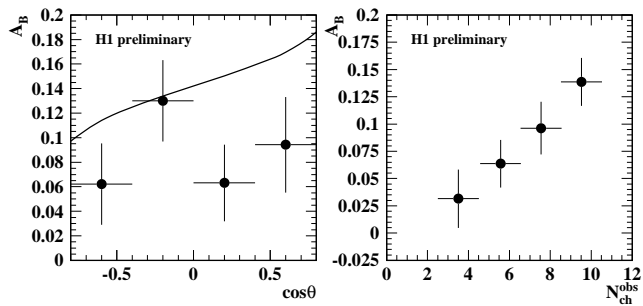}
          }
   \end{center}
\caption{\protect\label{fig-ab}
a) Baryon-antibaryon asymmetry $A_{\rm B}$ as a function of
$\cos\theta$ compared to the model of Ref.~23;
b) $A_{\rm B}$ as a function of the number of accompanying 
charged particles, $N_{\rm ch}^{\rm obs}$.}
\end{figure}
The $\cos\theta$ dependence of $A_{\rm B}$ is in qualitative
agreement with the data in Fig.~\protect\ref{fig-ab},
yielding a non-vanishing asymmetry at about 7 rapidity units
from the leading baryon production region. 
A quantitative
comparison is not yet possible, since the (anti-)proton momentum cut
and additional requirements on the multiplicity have
not be applied in the model.
In Fig.~\protect\ref{fig-ab}b, $A_{\rm B}$ is shown as
a function of the number of accompanying 
charged particles, $N_{\rm ch}^{\rm obs}$, within the same angular interval as
the (anti)-protons. The increase of $A_{\rm B}$ with $N_{\rm ch}^{\rm obs}$
is also in qualitative agreement with the expectation for
the gluonic mechanism of BN transport.
\section{Summary}
The L3 and OPAL measurements of the total hadronic
$\gg$ cross-section are consistent. The measurements
favour a stronger rise of the total cross-section with
energy than observed in $\gamma$p, pp or $\ppbar$ scattering.

The OPAL results on di-jet production in $\gg$ scattering
are well understood within the framework of perturbative NLO QCD.
The di-jet cross-sections help to constrain the gluon content
of the photon.

Comparisons of transverse momentum distributions 
measured in different processes yield information
about the partonic structure of the interactions.

A first H1 measurement of a non-vanishing baryon asymmetry 
in the photon hemisphere is in qualitative agreement with
the gluonic mechanism of baryon number transport.
\section*{Acknowledgements}
I want to thank Maria Kienzle (L3), Martin Erdmann, 
Andrey Rostovtsev and Martin Swart (all H1) for their help
in preparing this presentation.

\end{document}